# RATAN-600 and PLANCK Mission: new results after First General Meeting

Yu.Parijskij

Special Astrophysical Observatory, Russian Academy of Science

369167 Niznij Arkhyz, Karachay-Cherkessia, RUSSIA

## Abstract

We present activity at RATAN-600 connected with future PLANCK mission. It includes special radio telescope upgrading, multi-frequencies observations in the 0.6GHz-30GHz band with 40 channels, new results, connected with synchrotron, spinning dust and polarized Galaxy emission at small scales. Much lower limits were found than expected before the First Planck meeting. New deep sub-mJy survey of the 100 sq.deg. field at cm. waves demonstrated, that NVSS objects will dominate at LFI frequencies, and new, invisible by NVSS population will be below PLANCK sensitivity limit with only few exceptions. Preparation to the next generation experiments, connected with PLANCK mission we describe as well. They include high l polarization, SZ- blind survey, and test observations of the cosmological highly redshifted lines at small scales.

## I. Introduction

RATAN-600 [1], is the world biggest in size reflector, with resolution by two orders of magnitudes higher, than it is necessary just to resolve the CMB anisotropy. As any reflector, it has much higher brightness temperature sensitivity than any radio telescope with unfilled aperture. Excess resolution helps greatly in the filtration of the atmosphere noise [2],[3] and in the cleaning data from the point sources (PS). The most popular "Transit" mode of observation in the local meridian gives the unusual ability of multi-frequency observations at many wavelengths. Moreover, the aplanat solution with practically zero aberrations [4] may be realized near local zenith, and from hundreds to several thousands receivers array may be installed along the focal line of the secondary (parabolic cylinder) mirror.

Here we present the list of innovations of the RATAN-600 facilities, data accumulation process at the present state, results obtained up to now and activity connected with the next generation PLANCK mission related experiments at RATAN-600 [30].

II.RATAN-600 upgrading
a.       Panels resurfacing and optimization of the panels shape. To improve the efficiency at the highest frequency, close to PLANCK LFI, we resurfaced all 900 (7.5m x 2m in size ) elements of the RING part of the RATAN-600 and all 80 (8.3m x 3m in size) FLAT mirror elements. Now the panels have r.m.s. errors about 0.2mm., by factor 5 better, than it was before.
b.       Panels mean curvature was change to be optimized to the small zenith angle field, where two sources of systematic errors disappear (wrong curvature effects and wrong orientation effects, see [1].
c.       Panels adjustment process now under the full computer control and readjustment errors now are below the new surface accuracy.
d.       Two suppress the spillover effect, special screens (1.5m above and 1.5m below the main panel surface) were installed. It also increased the geometrical surface of the RING at the low frequency part of the RATAN-600 band by 5400 sq.m. and the lowest frequency of the RATAN-600 now moved to 300 GHz (Fresnel zone limitation).

e.     Main decemetrics frequencies (0.6GHz, 1GHz, 2.4GHz  with 8 sub-channels each) were integrated in the non-standard single phase center feed system (ref), which simplifies the interpretation of the Galactic backgrounds.

f.     Several innovations were tested to suppress the quickly increasing level of the man-made interference and some of them are in the regular use. Atmosphere interference was studied deeply [2], and limitation on the CMB anisotropy experiments was found in I, U, Q Stokes parameters. Most of the data  may be used for high l  scales (l>200-500) .

g.     Multi-frequency receiver array was improved  greatly: new 0.6GHz receiver was added and better crio radiometers appeared at high frequencies, and all broad band decemetrics receivers were subdivided by 8 channels. It increased the frequency channels number in 0.6GHz-21GHz range from 7 to 28. (see http://www.sao.ru)

h.     New experimental matrix HEMT  MMIC   receivers array at the PLANCK LFI frequency, 30GHz  was installed a (first version, with 6 horns), and new version is now in progress, with 32 horns. [5], [6].

i.     New two-channels 5GHz receiver was added recently,  which is the most sensitive now, and it will be used efficiently for the separation of the point sources from other backgrounds, as well as  for many calibration purposes [7].

Few important  software improvement and the simplification of the data access process may be mentioned  as well: quick "stream" data reduction algorithms appeared and optical fiber connection of RATAN-600 backend  with the main Observatory DATA BANK and INTERNET now available. Much better understanding of the  beam shape we have now, and improved theory was checked by observations accurately [8]. Very distant scattering, down to -70db was estimated now at all frequencies, and the level of this scattering is much below the PLANCK one. Muller matrix was calculated with real aberration effects, and it was shown that antenna instrumental polarization in U,Q   Stokes parameters is below 1% for   l<10000   CMB experiments [5].

### III. DATA accumulation  process.

About 1/3 of the telescope time last several years was spend on the PLANCK- related experiments . Before 2000 we had several CMB experiments (see [11]  and  references there), including COLD experiment concentrated on the DEC=5 deg.,  0h<RA<24h 1 degree wide strip survey. Later we added DEC=88 deg. 0h<RA<24h  strip and now the main region is connected with local zenith strip survey, DEC=41deg 20min  0h<RA<24h and 10 adjacent DEC strips, separated by 12.5 arcmin.

It was realized, that not only the CMB DEDICATED  projects data, but also all data accumulated in long-time monitoring programs (50% of the telescope time)  should be used for small scale anisotropy experiments. In the most popular TRANSIT multifrequency mode of observations,  all observers need to observe the field comparable with the scale of the cosmological horizon at the recombination epoch. About 1000 objects were observed more the 100 times, and high pixel sensitivity was accumulated in about 1000 sq. deg. total area. This approach reminds the 2dF one in  optical domain, and we call this type of our data flow as "2dF Radio". All observations are in the multi-frequency mode, with white noise sensitivity about few mK (1 sec. integration) at high frequencies (4GHz-30GHz ) and about 10mK-20mK (1 sec. integration) at low frequencies (0.6GHz-2.4GHz). The HPBW at 30GHz  is 0.1arcmin x 0.6arcmin in the ZENITH field, with non-zero extension up to 6arcmin in DEC. This beam is scaled with frequency in the standard way. Details of the RATAN-600 facilities may be found at the RATAN-600 site, http://www.sao.ru  [9], [10].

About  100 of  24-hours  scans collected at DEC=88 deg, 300-at DEC=+5 deg, 300 in ZENITH field and  more than 100 scans per each of the about 1000 2dF fields. Some part of the data are from single beam experiments, some from beam switching, and  at 7.7GHz and 30GHz  there is big collection of the U, Q data. Total data accumulation process at all high frequencies (4GHz, 7.7GHz, 11GHz, 21GHz, 30HGz) is visible on the Fig.1

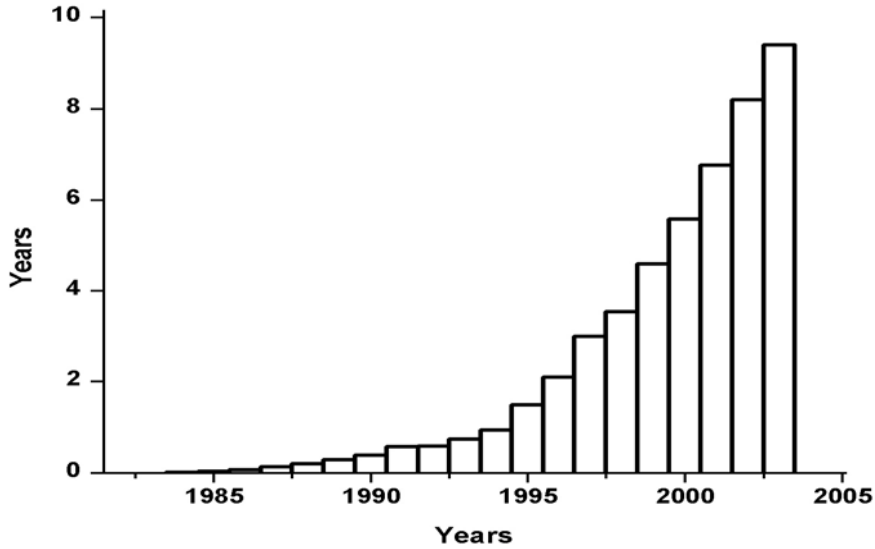

Fig.1. Total Telescope time, accumulated at RATAN-600 at 5 "Galaxy Window" frequencies in three types of programs: DEC=5 (Cold) field, ZENITH field , POLE field and "2dF Radio" (monitoring programs).

The main target- to accumulate pixel S/N ratio close to the CMB anisotropy level (or to the level of the secondary anisotropy effects). As far as the white noise receiver components, we are now close to the level, suggested by theory, and we recently changed the strategy, and began to extend the size of the ZENITH field, to avoid the "COSMIC VARIANCE" limitation. We hope to accumulate data as long as we need, may be, up to the PLANCK mission time (Feb.2007).

We have to compensate lower receivers NET sensitivity than WMAP, South Pole and PLANCK one by much longer integration time per pixel.

## IV. Some results.

a. Better understanding of the "Galaxy Window" to the CMB now available. Instead of HASLAM 408 MHz synchrotron template (too low frequency, too low resolution, 1K temperature resolution only) we spent few years to improve data on the CMB interesting l and frequencies, closer to PLANCK LFI than 408MHz HASLAM map case.

We realized, that the GALAXY window is much wider, than suggested by Early predictions at least for PESSIMISTIC and close to OPTIMISTIC case ([12]). Our new data [13], [14] are shown at the Fig.2.

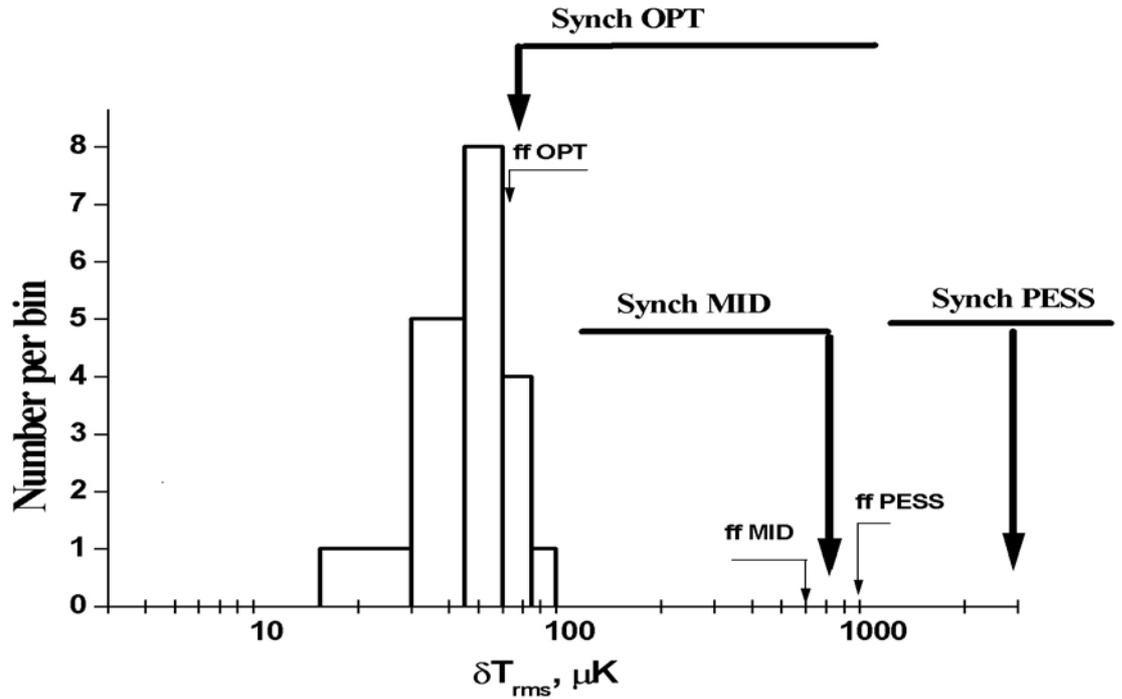

Fig2. Comparison of the old SREENS predictions for synchrotron, free-free noise at l=2000 with new RATAN-600 data at cm. waves. PESSIMISTIC, MIDDLE and OPTIMISTIC variants are shown from [12 ].  It is visible, that real noise is closer to the "optimistic" one.

b.      Two years we spent two years  to check the Princeton group prediction of  the noise from spinning dust [15] . We have frequencies, where this noise is the most intense, and found only the upper level of the spinning dust noise by order of magnitudes below predicted at CMB interesting scales (c.f. PESSIMISTIC case in [12]),  as visible from  Fig.3.

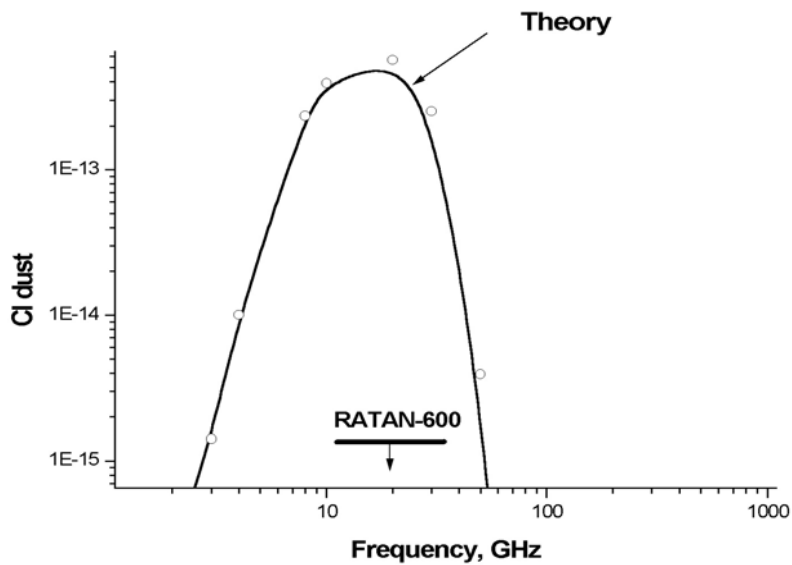

Fig.3. Comparison of the original "Spinning dust" theory [12, ] with RATAN-600 dedicated observations at predicted frequency,  21 GHz [17]. Our upper limit  is by order of magnitude below predictions.

c.     There were many estimates of the level synchrotron polarized noise from the Galaxy. Again, most suggestions were done using much lower frequencies,  less resolution, and less sensitive than it is necessary to predict accurately enough the l=1000 noise at PLANCK frequencies. We collected   about 1 year data at high Galactic latitudes at 7.7GHz and realized, that simple extrapolation in frequency, scales and latitudes [12], [21] does not fit our data- the real polarized noise is much less, and several explanations were given [13], [17].

d.     Point sources (PS) problem for PLANCK.
        About  five years ago, at the beginning of the PLANCK project activity,   there was discussion about the possible NEW PS POPULATION, connected with INVERTED, or even mm GPS objects,  DUSTY  objects invisible in Radio Catalogs, and about strong  SZ noise at very high l. Efficiency of the cleaning of the CMB experiments by RATAN-600 PS data was demonstrated in [29]

In January, 2004, we compiled the list of PS, detected in our deep zenith survey with at 4GHz with limiting flux density about 1mJy [18].

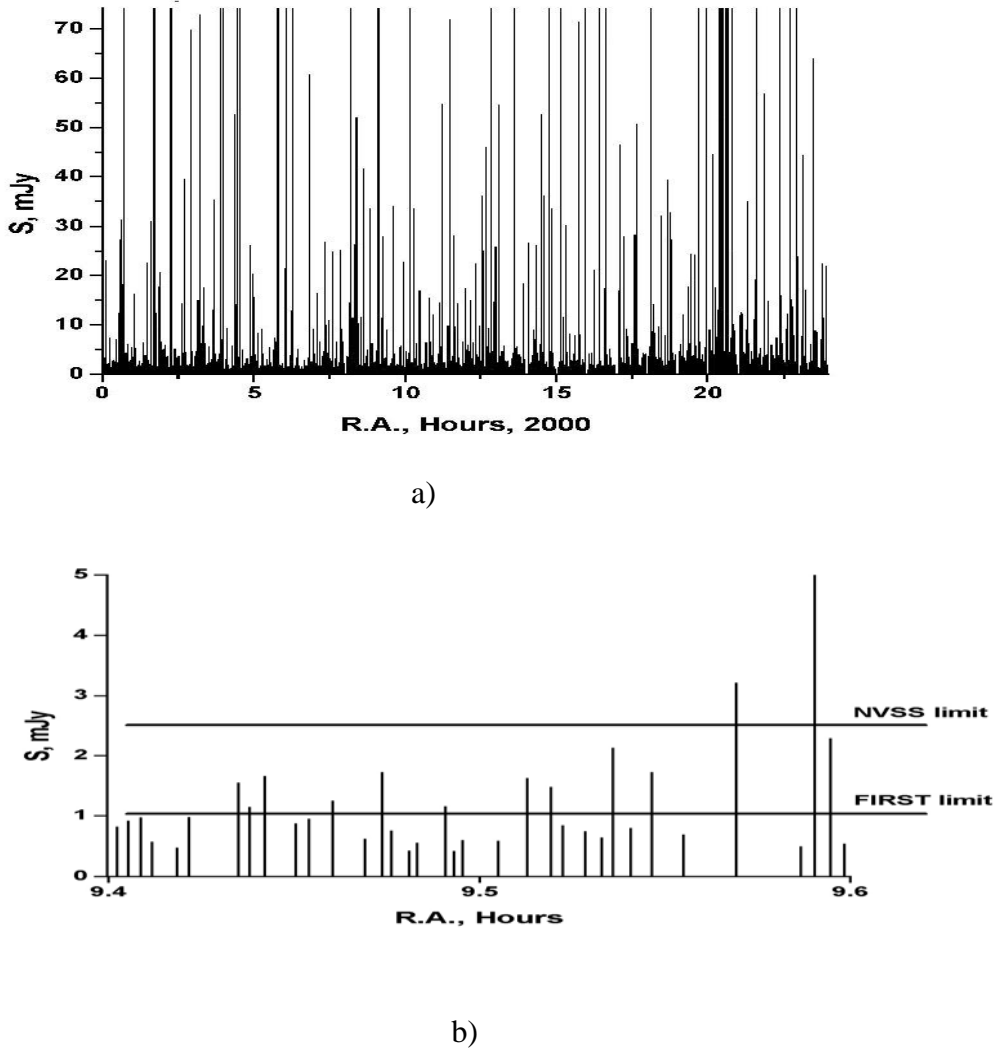

a)

b)

Fig.4. Deep strip scan result at 4GHz at DEC=41 grad. 20 min with selection of all details stronger than 2-sigma noise level [18]. These details (4452 in about 15 sq.deg. area) were separated from the scan before CMB  and background anisotropy estimates  were made. a)- all details in the 0h<RA<24h,  DEC=41 grad. 20arcin +/-1.5 arcmin  strip area; b)- small part of the scan between  9.4h<RA<9.6h are shown.

The surface density of these sources happened to be close to that of the best decemetrics NVSS and FIRST catalogs and much greater that in any cm. waves catalogs, based on the observations of sky area greater than 1 square degree, see Fig.5 below, were only strong (more than 5-sigma noise level objects were counted.

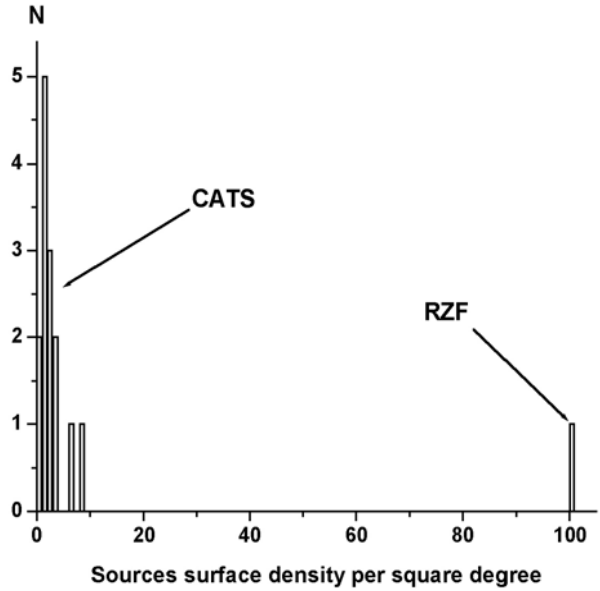

Fig.5. RATAN-600 Zenith Field (RZF) strong source population surface density per square degree compared with that for the best cm-waves catalogs compiled from the sky area greater than 1 sq. deg. list (see CATS, http://cats.sao.ru, e.g.)

Practically all listed objects were never observed in cm. band, 70% may be identified with NVSS objects, about 30% are new radio sky objects.

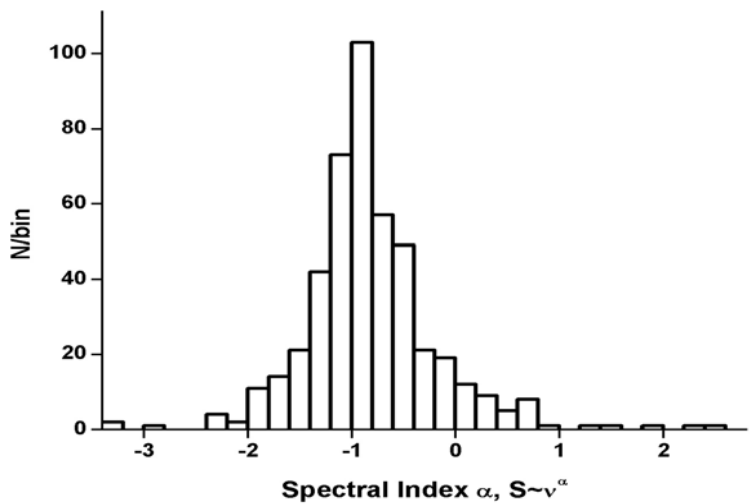

Fig.6. Simple 2-frequencies spectral indexes histogram of NVSS objects, found in our ZENITH strip survey. Median $\alpha$=-0.85, with few USS and inverted spectra objects.

. From dm-cm spectral indexes (see Fig.6) we have tried to estimate, how many NVSS objects may appear in PLANCK maps (say, above 300 mJy at 30GHz ) and realized, that this number is small. All new objects are weak, with median flux density at 4GHz about 2mJy, and even in the extreme cases of inverted , GPS, Black body spectrum this new population can't be dominant one at PLANCK LFI frequencies. First data reduction of 11GHz data in the same zenith strip survey confirms this conclusion. We suggest, that most objects of the PLANCK LFI Sky will be NVSS objects, with some mixture with IRAS objects.

Immediately after publication of 208 WMAP objects, with 8 new sky objects, we compiled the spectra of all identified objects and reobserved some of them 19] . We do not confirm the reality of NEW sky objects . At the same time, it was demonstrated, that real nature of objects may be better understood if spectra compilation will be based on the observations made as close in time to the PLANCK observations, as possible. In cooperation with Finland PLANCK group, we are preparing now to the "Quick Reaction" mode of observations, were all 8 broad band RATAN-600 (with about 40 sub-channels) may be used for compilation of spectrum in the INTERDAY time scale. The first experience with WMAP PS list suggests, that no less informative may be just coordination of the global monitoring programs, conducted now by several groups with different scientific accents, with PLANCK. With predicted BEAM SCANNING process, it is easy to prepare the "time table" for the moments, when PLANCK beams cross the most interesting sky objects. PLANCK + HERSHEL common field may be included in this list as well.

e. Jupiter as a PLANCK calibrator.

We were asked by TAC PLANCK group to check the stability of Jupiter radiation at 30GHz, and how accurately it may be used as a calibrator during PLANCK mission. About hundred 6 arcsec resolution in R.A. and with S/N =1000 scans at 30GHz demonstrated, that the mean disk temperature is stable enough, (at –30db level). Belt emission also was barely detected, at the level about or less than 1%, but this emission is also stable enough (at <10% level) [20]

## V. RATAN-600 next generation experiments.

Some directions of pre-PLANCK activity in the next 4 Years are well formulated and are in progress.

a. First of all it is an escalation of the rate of data accumulation by increasing of number of receivers used. Main road- 30GHz matrix array. 6 horns array now in real observations, and 32 horns in the construction, we hope test it in the next winter.

256 horns matrix is waiting for the financial support, but special big secondary mirror with 4.8m focal line is ready. New much better rail tracks were constructed recently which can guide this new 150 ton (7.4m x 11.5m ) parabolic cylinder. Instead of standard speed (square root of the telescope time, T), the square root of the product T x N(t) is now the main parameter, and 32 horns array supply result of 1 month averaging of single horn observations in one day.

We expect the rate of the data accumulation close to that, shown in fig.7. With single cannel in the transit mode of observations the pixel sensitivity in one scan only is about 0.4 mK for l=2000 scale. Accumulation of the 1000 scans (1000 days x number of channels) gives much better sensitivity.

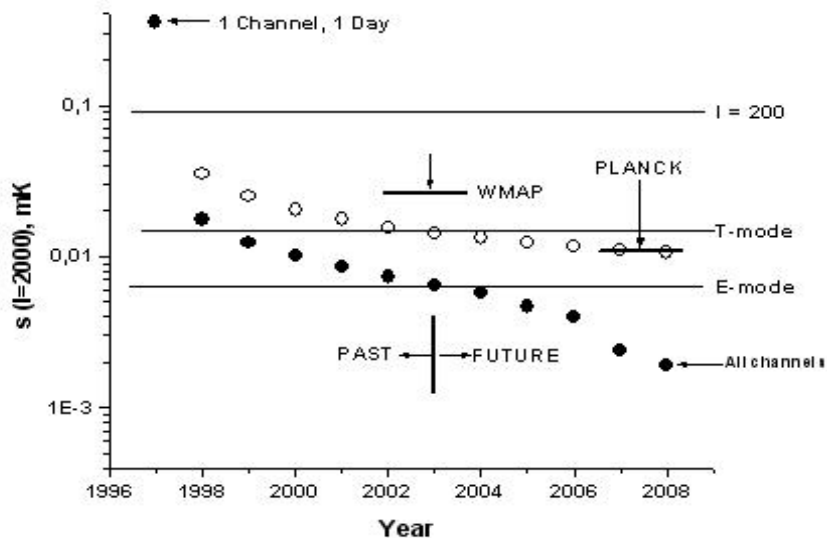

Fig.7. Pixel sensitivity accumulation process at l=2000 scale. Open circles- one channel sensitivity, filled circles- sum of all channels in the "Galaxy Window" band (f>4GHz). S/N=1 levels are shown for T and E CMB anisotropy modes (horizontal lines); l=200 level is shown as well.

b.    High l polarization.

With 30GHz matrix we begin to search small scale sky polarization, and in the limited (say, 100 sq. deg. ) field to be close (or better) than PLANCK mission suggests, and some information we expect to have up to l=5000. With S/N>1 in the small region the "Cosmic Variance" noise will dominate, and with no resolution limitation we can expect the result shown in Fig.8.

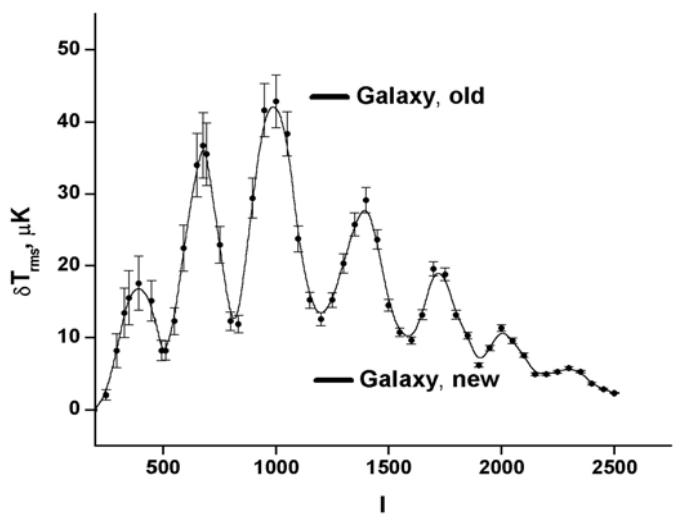

Fig.8. Expected "Cosmic Variance" noise in E-mode of CMB polarization after accumulation of the data in the ZENITH Field at 30GHz up to S/N <1, [31]. Old [12] and RATAN-600 new estimates [17,18 ] of the Galaxy Synchrotron noise at high l are shown by horizontal bars.

c.    SZ Cosmology

Simple Thermal SZ effect (first observed in 1970, Pulkovo, with RATAN-600 type 130m-Radio Telescope, [22]) should dominate in the sky noise at l>3000. The present day pre-PLANCK activity in this field connected with mapping of known X-Ray clusters. But the X-Ray emission quickly escapes from the detection at z>1-2, whereas the effect itself does not depend on the redshift, and may exist as early as z=5-30. That is why the "blind SZ survey" is sharply needed. We suggest RATAN-600 as a tool for such observations. Flat mirror inside the circle in combination with southern sector (see www) may be used as blind survey instrument in simple transit mode with big daily field of view. With new generation cooled array with sub-mK NET, we can detect hundreds SZ spots in the single day. The result of simulation of such experiments with present day logN-logS data is shown at Fig.9 [5].

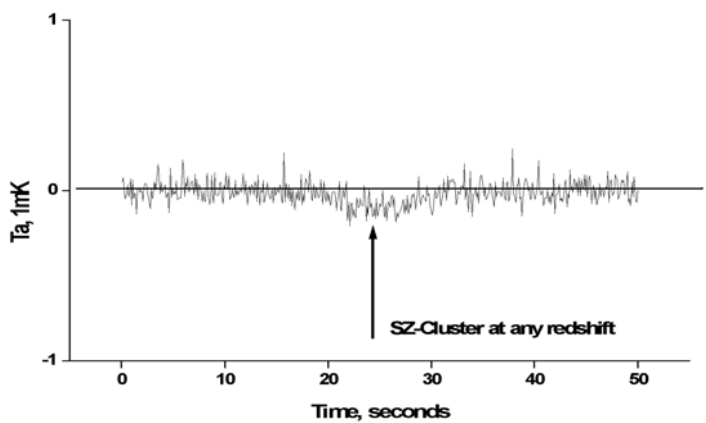

Fig.9. Transit curve of standard distant X-Ray cluster at RATAN-600 using with FLAT mirror mode (frequency 30GHz, 6 arcsec x 6 arcmin resolution, 0.5 seconds integration time). PLANCK LFI type receivers array along the focal line of the secondary parabolic cylinder mirror were used in simulation. Point sources may be easy separated due to high resolution.

They can be studied accurately with dedicated SZ instruments. Few arcsec (one-dimensional) resolution results in good position and angular size measurements, and RJ specific ("Black body with negative temperature" spectrum may be confirmed at RATAN-600 or elsewhere. "Sign, Size, Spectrum" method in RJ band leaves no room for other interpretation. After simple detection much greater S/N data should be collected, and for this purpose we are preparing to the few hours integration mode, using special rail track inserted between FLAT mirror and SOUTHERN sector. This track was fully reconstructed recently to meet accuracy specified by 30GHz frequency. There is also some very distant variant to transform the simple total power receivers array into the "phased focal array". In this case it is possible to correct aberrations of all kinds, increase greatly the daily field of view and convert single aperture mode of observations into the "multi-sub-aperture" system, with sub-aperture size as small as size of telescope panel, [23]-[25], [32].

CMB Spectroscopy belong to the most difficult but very interesting field of cosmology, where only few theoretical publications available [27] and practically no experiments ( except some preliminary test by Gosachinskij group. Reflector geometry is the only one which may be used in "blind" searches of spectral sky features at unknown redshift (say, 5<Z<50; 700<Z<3000). HI, H2O, CO, primordial moleculas, recombination lines from recombination and secondary ionization epoch- among the candidates. Computer control of the telescope panels positions gives also unusual possibility of Michelson Fourier interferometry, but it was never tested at RATAN-600 [23]). RATAN-600 High resolution we use in the test CMB spectroscopy observations to find any "Frequency-Space structure" and first results were published recently.

We are very interested in cooperation; it makes our long distant plans more reliable.

## Acknowledgements

A large number of staff members of Special Astrophysical Institute and several others institutes took part in the PLANCK ground based support activity with the help of RATAN-600. Improvements of the telescope facilities were done by Zhekanis G.V., Khaikin V.B., Golosova , Zverev Yu.K. Majorova E.K., matrix receivers array was developed by Berlin A.B, Nizhelskij N.A. groups in cooperation with SATURN (Ukraina), Observations were done mainly by Bursov N.N. under general supervision by Mingaliev M.G. Chepurnov A. , Verkhodanov O.V. Kononov V.K.were connected with software support. PLANCK TAC group in Denmark (Novikov I.D. et al) were important in the selection of the problems to be solved with RATAN-600 facilities. We used RFBR grant 02-02-17819, ASTRONOMY, "INTEGRACIA" grants and special Saint-Petersburg Scientific Center of the Academy of Science grant.